# Ferromagnetism in CaRuO$_3$ thin films by efficient route of tensile epitaxial strain

Shivendra Tripathi, Rakesh Rana, Parul Pandey, R.S. Singh and D.S. Rana*

*Indian Institute of Science Education and Research (IISER) Bhopal, M.P. – 462023, INDIA*

**Abstract**

We show that a ferromagnetic (FM) order in the orthorhombic CaRuO$_3$, which is a non-magnetic and iso-structural analog of FM system SrRuO$_3$, can be established and stabilized by the means of tensile epitaxial strain. Investigations on the structural and magnetic property correlations in the CaRuO$_3$ films with different degrees of strain reveal that the FM moment increases with increasing the tensile strain. This is an experimental verification to the theoretical predictions of scaling of the tensile epitaxial strain and the magnetic order in this system. Our studies further establish that the tensile strain is more efficient than the chemical route to induce the FM order in CaRuO$_3$ as the magnetic moment in these strained films is larger than that in chemically modified CaRu$_{0.9}$Cr$_{0.1}$O$_3$ films.

* Email: dsrana@iiserb.ac.in

Among various 4d transition metal oxides, metallic $SrRuO_3$ is the only system which exhibits a long range ferromagnetic order with a Curie temperature of 165 K.[1-3] Large magnetic moment and the metallic character of $SrRuO_3$ make it as one of the most suitable material as ferromagnetic metal electrode in the spintronic devices based on spin-polarized tunnel junctions.[4] Due to various fundamental and technological interests in $SrRuO_3$, it is only logical to look for analogous systems in this class of ruthenates exhibiting intriguing properties. In this regard, $CaRuO_3$ is a natural choice as it is iso-structural and iso-electronic with $SrRuO_3$ and is metallic down to low temperatures.[5] It also exhibits a wide variety of properties such as non-fermi liquid behaviour,[5] magnetic quantum criticality,[6] pressure induced post-perovskite structure,[7] etc. However, contrary to expectations the $CaRuO_3$ does not exhibit a long-range magnetic ordering. Known to be a metallic paramagnet,[8,9] some studies indicate the $CaRuO_3$ to exhibit antiferromagnetic order with $T_N \sim 110$ K.[2] Its magnetic state, however, is still debated and is far from established. One agreement that has been reached among various researchers is that the $CaRuO_3$ is on the verge of establishing correlations to account for a magnetic order at low temperatures.[2,10] It is understood that in $ABO_3$ structure the difference in ionic radius of A-site cations $Sr^{2+}$ (~1.31 Å) and $Ca^{2+}$ (~1.12 Å) is responsible for the difference in the ground state of $SrRuO_3$ and $CaRuO_3$.[11] Owing to this reason, there is a definite interest in understanding and manipulating the magnetic ground state of $CaRuO_3$ by the means of chemical substitution, disorder and epitaxial with an aim to obtain ferromagnetic order as in its counterpart $SrRuO_3$.

Several studies have suggested the metallic $CaRuO_3$ to be on the verge of establishing ferromagnetic phase transition.[6,12-14] The $CaRuO_3$ is a metal with a $GdFeO_3$ type orthorhombic structure (a = 5.541 A, b = 5.362 A, and c = 7.686 A, space group – Pnma). The central Ca atom is surrounded by corner sharing $RuO_6$ octahedra.[15] The distortion of $RuO_6$ octahedra affects the Ru-O-Ru bond angles, which consequently affects the electronic and magnetic properties.[11] A magnetic order in $CaRuO_3$ may be expected if the Ru-O-Ru bond angles and bond distances can be manipulated by two primary means, namely, chemical means of A and B site substitution or physical means of epitaxial and uniaxial strains in thin films. He and Cava reported that disorder created by mere 2% doping of non-magnetic Ti at Ru site in $CaRuO_3$ induces ferromagnetism in the system.[6] Extensive studies on $CaRu_{1-x}M_xO_3$, where M (transition metal) is either a magnetic or non-magnetic ion, have shown doping induced discernible modifications of the magnetic and electronic phases.[16,17] The Cr substitution for Ru have proved quite effective as it induces substantial ferromagnetism in CaRuO$_3$.[18-20] Resultantly, the polycrystals of Cr-doped CaRuO$_3$ have been the subject of many meticulous studies in recent times.[18,19] A detailed study revealed that Cr substitution as low as 2.5 % induces ferromagnetism in CaRuO$_3$ and that the magnitude of magnetic moment increases with increasing Cr content, reaching the maximum value for 15% Cr substitution. This trend reverses as the Cr concentration exceeds 20%.[18]

Understanding and manipulating the magnetic ground state of metallic CaRuO$_3$ is one of the key issues in perovskite ruthanates. All above-mentioned methods to induce ferromagnetic moment in CaRuO$_3$ involve chemical substitution that disrupts the Ru sub-lattice. Zayak *et al* showed theoretically that by applying tensile strain on CaRuO$_3$ results in a transition from a nonmagnetic to ferromagnetic (FM) state and that the magnitude of induced FM moments scales with the tensile strain.[21] Compressive strain, on the other hand, does not modify the magnetic state. However, an experimental study in which FM moment is induced as a function of tensile epitaxial strain in single layer phase-pure films is not yet realized and established. We have deposited the pure and 10% Cr-doped CaRuO$_3$ thin films on substrates with lattice constants inducing either the compressive strain or the tensile strain. In this letter, we show that the tensile strain for films on SrTiO$_3$ (100) substrate induces a weak FM order for pure CaRuO$_3$. The magnetic moment scales with the tensile strain, which is commensurate with the theoretical predictions. Furthermore we also demonstrate that the tensile strained CaRuO$_3$ films possess larger magnetic moment compared to that of chemically modified CaRu$_{0.9}$Cr$_{0.1}$O$_3$ films.

Polycrystalline samples of CaRuO$_3$ and CaRu$_{0.9}$Cr$_{0.1}$O$_3$ were prepared by standard solid-state reaction route. The x-ray diffraction data confirmed the phase purity of both the samples. These samples were used for preparation of thin films using pulsed laser deposition technique. Films of various thicknesses, in the range of 20 – 140 nm, were deposited using a 248 nm KrF excimer laser. The parameters for various depositions were: energy density between 1.7-3.3 J/cm$^3$, laser pulse frequency - 4 Hz, substrate temperature - 700 °C, O$_2$ partial pressure - 40 Pa, O$_2$ annealing pressure of 1000-1800 Pa. The SrTiO$_3$ (100) substrate (lattice constant ~ 3.905Å) with a mismatch of about 2 % was chosen to obtain tensile strained CaRuO$_3$ thin films. Though phase-pure oriented films were formed on BaTiO$_3$ (100) [a ~ 3.99 Å] and MgAl$_2$O$_4$ (100) [a ~ 4.04 Å] substrates, strained films could not be obtained as the lattice mismatch of these substrates with CRO is too large to be accommodated for stability of strained phase. Only films on SrTiO$_3$ (100) substrate could be stabilized with a reasonable tensile strain. It is known that the CRO films on SrTiO$_3$ (100) have a tendency for formation of pseudo-heterostructures.[22] The problem with these films is that it is difficult to assign the origin of magnetic moment, if any, to any of the co-existing phases. Hence, it is required to segregate these epitaxial phases and investigate their magnetic properties. We started the usual deposition by varying the energy density and keeping other parameters fixed. The film obtained with energy density of 1.7 J/cm$^2$, say CRO-A, possessed two phases as evident from two closely spaced epitaxial reflections in θ-2θ patterns (Figure 1a). To get rid of one phase in this pseudo-heterostructure, the energy density was increased to 2.3 J/cm$^2$. Thus obtained film, say CRO-B, too possessed two epitaxial phases. Finally, films with single homogenous phase were obtained when laser energy density was fixed at 2.0 J/cm$^2$. With these optimized parameters, films with thickness of 140 nm and 31 nm, respectively, labeled as CRO-C and CRO-D were deposited. For $CaRu_{0.9}Cr_{0.1}O_3$, the 130 nm and 30 nm films ($CRO_{10}$-A and $CRO_{10}$-B, respectively) were deposited with optimized energy density of 2.0 J/cm$^3$.

The θ-2θ x-ray diffraction patterns acquired on 'Emperean' PanAnalytical x-ray diffractometer showed all films to be phase-pure. The ω-2θ reciprocal space maps (RSM) using a four axis cradle mounted on same diffractometer were also obtained for detailed structural and strain analyses of the films. The resistivity and magnetoresistance measurements were performed using the four-probe technique in a temperature range 2K-300 K. The magnetization measurements were performed using superconducting quantum interference device magnetometer.

Figure 1 shows the structural properties of all the CaRuO$_3$ films. In fig 1(a), the θ-2θ patterns of four representative structures are shown. The CRO-A and CRO-B films possess two structures. There is a clear splitting of both (100) and (200) peaks in CRO-A film (fig 1a), which arises from co-existing structural polymorphs, namely, fully and/or partial strained orthorhombic and cubic phases of the CRO film. On the other hand, the optimized films, namely, CRO-C and CRO-D showed only one epitaxial reflection suggesting these films existing in only one structural form. Detailed structural investigations of structure were carried out by acquiring the RSM around asymmetric (301) and (311) peaks. Figure 1 (b-d) shows RSM for CRO-A, CRO-C and CRO-D films around (301) peak. The salient features of these data are: i) similar to that of data in fig 1(a), these RSM data also show that films CRO-A and CRO-B shows dual peaks while the remaining show only one epitaxial peak, ii) the peak of the films do not lie on the pseudomorphic line of the substrate STO implying that none of these films is coherently strained and that the in-plane lattice parameters of the substrate and film are not the same, iii) among the films exhibiting only one peak (i.e., in CRO-C and CRO-D), the out-of plane lattice parameter of 3.88 Å for CRO-C films with a thickness of 130 nm decreased to 3.86 Å for CRO-D film having a thickness of 31 nm. This suggests that the CRO-D film possesses larger tensile strain than the CRO-C film. This was supported by the variations in the in-plane parameters of all these films, as calculated from the RSM data. One would expect that in strained films, as out-of-plane lattice parameters decreases, the in-plane parameter would increase to accommodate the tensile strain. We observed the same as one of the in-plane lattice constant increased from 3.87 Å for CRO-C film to 3.89 Å for CRO-D film. In CRO-A and CRO-B films, the out-of-plane lattice constant of either of the reflections do not match with that of CRO-C or CRO-D film. In-plane parameter of one of the reflection, however, matches with that of CRO-C film. Overall, these data suggest that all the films are partially strained film, among which the films with single structural phase exhibit increasing tensile strain with decreasing thickness.

Figure 2 (a and b) shows the zero-field-cooled (ZFC) and field-cooled (FC) magnetization (M) versus temperature (T) data obtained in a field of 500 Oe for all the $CaRuO_3$ films. It is seen that the films with only one phase, *i.e.*, CRO-C and CRO-D, exhibit ferromagnetic (FM) like phase transition whereas CRO-A and CRO-B films possess the paramagnetic phase down to low temperatures. In the CRO-C film, the transition temperature is not well defined, but the bifurcation between ZFC and FC curves clearly points towards weak FM phase in this film. The CRO-D film, however, exhibits a pronounced magnetic transition with clearly discernible transition temperature in the vicinity of 100 K. Also, the FC magnetic moment of this film is about 3-4 times more than that of CRO-C film. This suggests that magnetic transition that tends to set in thicker film (CRO-C) manifests itself more clearly in more strained CRO-D film. To ascertain the occurrence of the FM phase in these films more firmly, magnetization versus magnetic field isotherms for all the films were also collected [Fig 2c]. A lack of magnetic hysteresis and linear variation of magnetization with magnetic field for CRO-A and CRO-B films unambiguously confirmed these films to be of non-magnetic nature. The CRO-C and CRO-D films, whereas, showed a FM like hysteresis in magnetic isotherms which clearly corroborated their temperature dependent magnetization data. Furthermore, the saturation magnetic moment of 31 nm CRO-D film is significantly larger than that of 140 nm CRO-C film. From both the temperature- and field-dependent magnetization data of CRO-C and CRO-D films, it may be inferred that the magnitude of magnetic moment scales with the tensile strain.

The structural and magnetic property correlations were investigated in weakly FM $CaRu_{0.9}Cr_{0.1}O_3$ films, namely, a 30 nm thin CRO10-A film and a 130 nm CRO10-B film. Figure 3 summarizes the XRD and the temperature- and magnetic field-dependent data of these films. Similar to that in pure CRO films, the out-of-plane lattice constant decreases from 3.89 Å to 3.87 Å as the film thickness decreases from 130 nm for CRO10-B film to 30 nm for CRO10-A film. A consequent increase in in-plane lattice parameters, as confirmed by RSM data (not shown here), suggests a larger tensile strain for 30 nm film. The ZFC-FC magnetization versus temperature data shows that both the films exhibit FM transition at around same temperature (fig 3b). The magnetization-field isotherms clearly show that the saturation magnetization increases with decreasing thickness of the film, which implies that the FM moment increases as the tensile increases in Cr-doped films (fig 3 c). There are two implications of these results, namely, i) the scaling of tensile strain and magnetization is similar to that observed for pure CRO films and ii) magnetic moment of pure strained CRO-D film is marginally large compared to that of Cr-doped strained $CRO_{10}$-A film. This is surprising because Cr-doping alone has proved most efficient in inducing the FM order in CRO.[18] In present case, the Cr-doped films were strained to same extent as the pure films. Despite this, larger moment in pure films suggests that the tensile epitaxial strain is clearly more efficient than the chemical route to induce magnetic order in otherwise non-magnetic $CaRuO_3$.

It is known that the quality films on STO exhibit bulk like metallic behavior whereas the deterioration in quality results in semiconducting behavior.[23] The temperature dependence of electrical resistivity for all the pure and Cr-doped $CaRuO_3$ films is plotted in figure 4. A decrease in resistivity with decreasing temperature for all un-doped films suggests metallic behaviour. The Cr-doped films, however, exhibit semiconducting behaviour (inset fig 4). In both the cases, the resistive behaviour is representative of their respective polycrystal bulk counterparts.[2] These data corroborate our assertions on the quality of films as indicated by the structural and magnetization data.

For over more than a decade, there have been several studies, mostly theoretical, on understanding the magnetic ground state of $CaRuO_3$.[10,11,21,24-28] This is mainly because in bulk form it lacks any FM order whereas its magnetic analog $SrRuO_3$ exhibits long range FM order. In several theoretical studies it was predicted that by any suitable means, chemical or physical, if the tilt and rotation of $RuO_6$ octahedra can be reduced and Ru-O-Ru bond distances and angles increased to move towards those in $SrRuO_3$, a FM order might set in $CaRuO_3$.[11,25] A manifestation of same was unambiguously observed by chemical means, via partial substitution of Ru both by magnetic ions as $Cr^{18}$ or $Fe^{29}$ and by non-magnetic ion as $Ti^6$. Neaton *et al* carried out detailed calculations of correlations of the epitaxial strain and magnetic moment in $CaRuO_3$.[21] A ferromagnetic $CaRuO_3$ was predicted to be formed by the means of inducing the tensile strain and in which the FM magnetic moment increases with increasing tensile strain. Compressive strain, whereas, would not yield the same effect. An increase in-plane Ru-O-Ru bond distances and a decrease in covalent character are two essential factors to induce magnetic order in $CaRuO_3$ and this can be achieved by tensile strain. It was shown that a tensile strain of 2% can induce a saturation magnetic moment of about 0.5 $\mu_B$/f.u. In the present study, we find the experimental evidence to these theoretical predictions. In our study, it is clearly seen that a tensile strain of about 1% in CRO-C film induces a magnetic moment of ~0.1 $\mu_B$/f.u. and an enhanced strain of about 1.5% in CRO-D films results in a magnetic moment ~0.3 $\mu_B$/f.u. We could not succeed to form $CaRuO_3$ films with even higher tensile strain on substrate such as BaTiO3 (mismatch ~ 4%) and $MgAl_2O_4$ (mismatch ~ 6%). It may be noted that the indications of FM moment were found in pseudo-heterostructures of $CaRuO_3$.[22] However, in such cases it is difficult to assign the occurrence of magnetic moment to a particular phase. In this case, a small magnetic moment was attributed to the cubic phase coexisting with relaxed and coherent orthorhombic phases. In comparison to this, the present studies show manifestation of an FM order with a clear magnetic transition in tensile strained orthorhombic $CaRuO_3$ single phase films.

In summary, we have fabricated $CaRuO_3$ films with single structural form and with varying tensile strain. We provide explicit proof to the theoretical predictions that FM moment in $CaRuO_3$ can be induced by the means of tensile strain and that the magnitude of magnetic moment increases with the tensile strain. We further show that tensile strain is more efficient than chemical route to induce magnetic order in $CaRuO_3$. These leaves an intriguing aspect open: is it possible to induce larger magnetic moment of up to 1 $\mu_B$/f.u in $CaRuO_3$ by adopting a combined approach of tensile epitaxial strain and optimal chemical substitutions?

This work was supported by the Department of Science and Technology (DST), New Delhi under the research Project No. SR/S2/LOP-13/2010.

Figure 1:

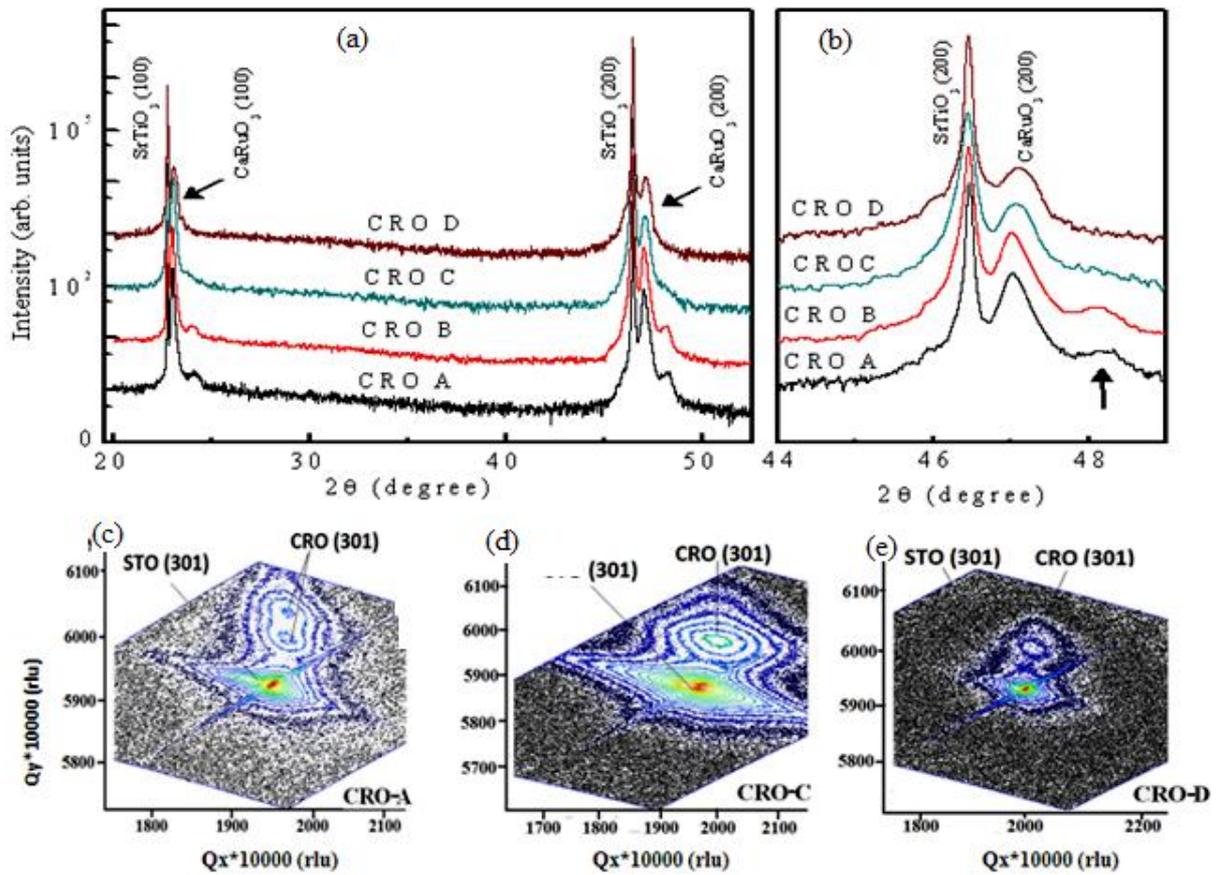

Figure 1: (a) θ-2θ XRD patterns for CRO-A, CRO-B, CRO-C and CRO-D films; (b) (200) peak is magnified in which the arrow indicates another CRO phase. Reciprocal space maps for (c) CRO-A film, (d) CRO-C film and (d) CRO-D film.

Figure 2:

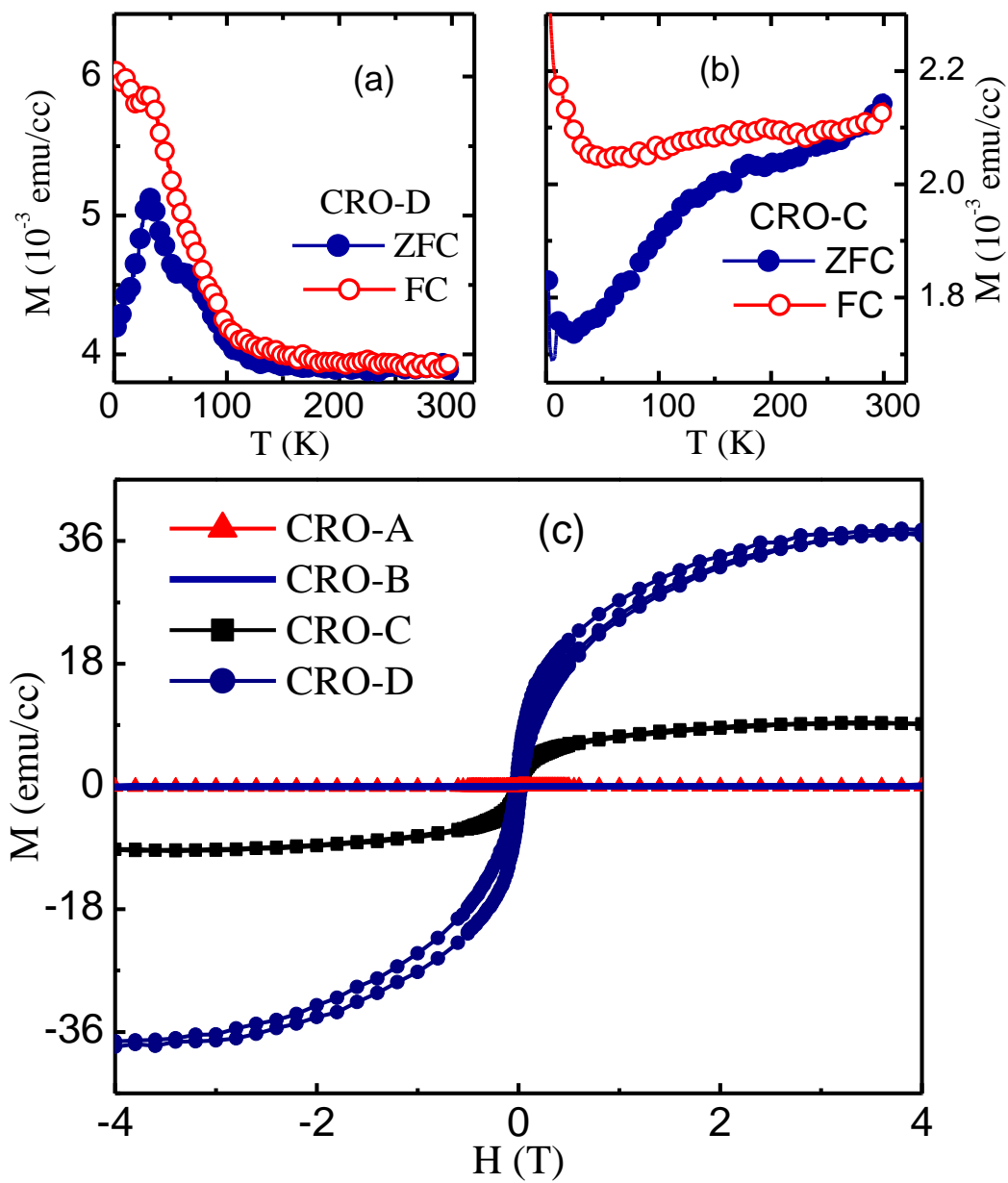

Figure 2: Zero-field-cooled (ZFC) and field-cooled (FC) Magnetization (M) versus temperature plots for (a) CRO-D film, and (b) CRO-C film. (c) Magnetization (M) versus field (H) isotherms for CRO-A, CRO-B, CRO-C and CRO-D films.

Figure 3:

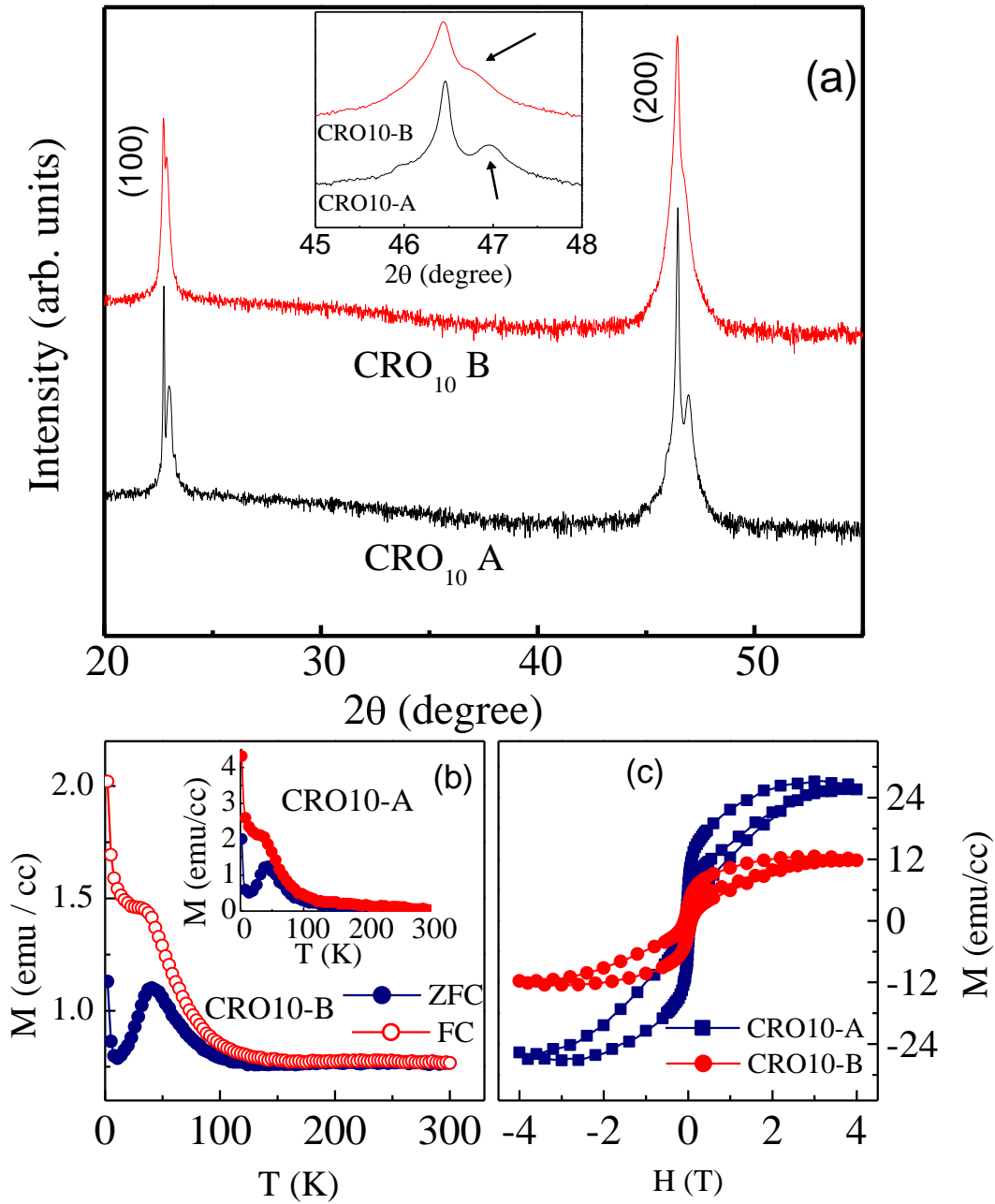

Figure 3: (a) θ-2θ XRD plots for CRO10-A and CRO10-B films. The (200) peak is magnified in the inset figure. Arrows in this figure indicate the (200) peak of $CaRuO_3$. (b) Zero-field-cooled (ZFC) and field-cooled (FC) Magnetization (M) versus temperature plots for CRO10-A and CRO10-B films. (c) Magnetization versus field (H) isotherms for same films.

Figure 4:

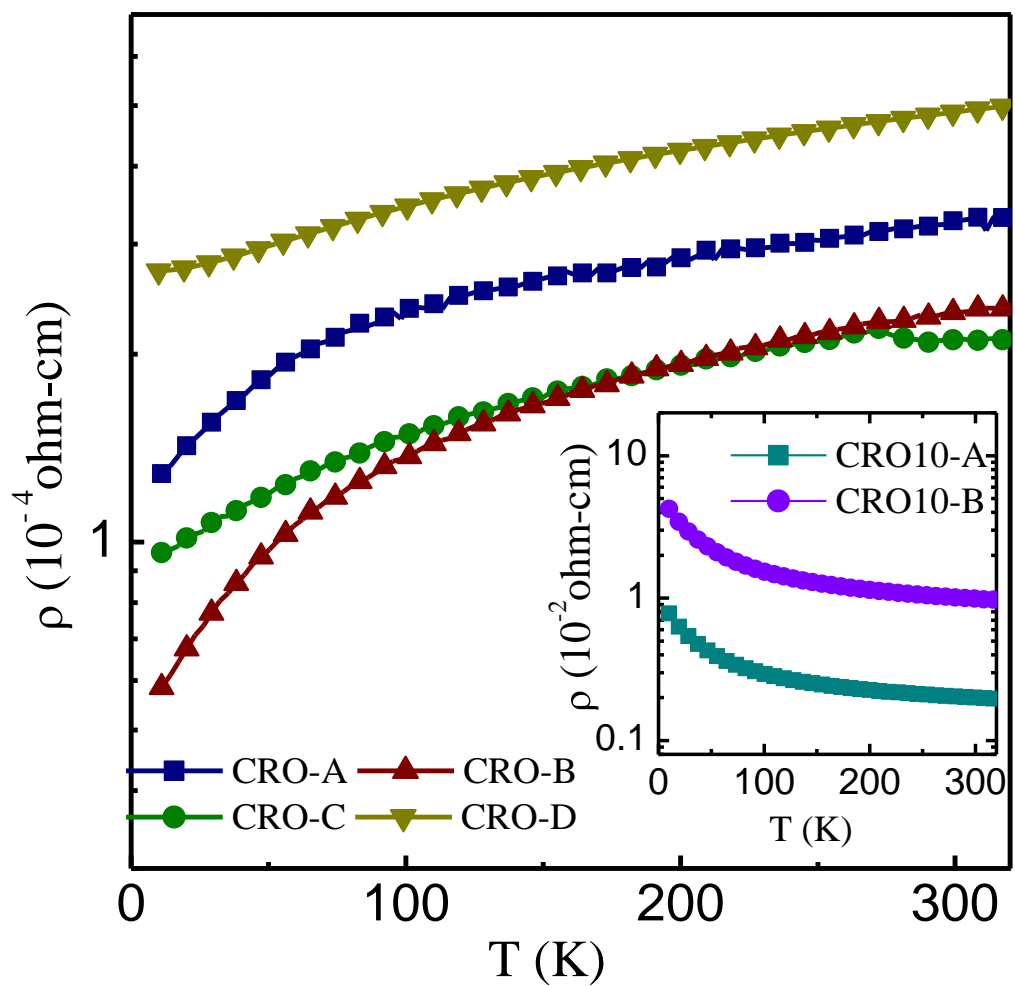

Figure 4: Resistivity (ρ) as a function of temperature (T) data for all the pure $CaRuO_3$ films (main panel) and Cr-doped films (inset).